\documentclass[a4paper]{article}

\usepackage{icrc2013}
\usepackage[english]{babel}

\usepackage[none]{hyphenat}

\title{The dual-mirror Small Size Telescope for the Cherenkov Telescope Array}

\shorttitle{The dual mirror Small Size Telescope for CTA}

\authors{
G. Pareschi$^{1}$, 
G. Agnetta$^{1}$,  
L. A. Antonelli$^{1}$, 
D. Bastieri$^{1}$, 
G. Bellassai$^{1}$,  
M. Belluso$^{1}$, 
S. Billotta$^{1}$,  
B. Biondo$^{1}$, 
G. Bonanno$^{1}$,   
G. Bonnoli$^{1}$
% ...
for the CTA collaborations 
}
\afiliations{
$^1$ INAF \& the ASTRI collaboration\\
$^{2}$ Paris Observatory \& the GATE collaboration\\
$^{3}$ The CHEC collaboration, various affiliations\\
}

\email{giovanni.pareschi@brera.inaf.it}

\abstract{ In this paper, the development of the dual mirror Small Size Telescopes (SST) for the Cherenkov Telescope Array (CTA) is reviewed. Up to 70 SST, with a primary mirror diameter of $\sim4$ m, will be produced and installed at the CTA southern site. These will allow investigation of the gamma-ray sky at the highest energies accessible to CTA, in the range from about 1 TeV to 300 TeV. The telescope presented in this contribution is characterized by two major innovations: the use of a dual mirror Schwarzschild-Couder configuration and of an innovative camera using as sensors either multi-anode photomultipliers (MAPM) or silicon photomultipliers (SiPM). The reduced plate-scale of the telescope, achieved with the dual-mirror optics, allows the camera to be compact ($\sim 40$ cm in diameter), and low-cost. The camera, which has about 2000 pixels of size 6$\times$6 mm$^2$, covers a field of view of $\sim10^\circ$. The dual mirror telescopes and their cameras are being developed by three consortia, ASTRI (Astrofisica con Specchi a Tecnologia Replicante Italiana, Italy/INAF), GATE (Gamma-ray Telescope Elements, France/Paris Observ.) and CHEC (Compact High Energy Camera, universities in UK, US and Japan) which are merging their efforts in order to finalize an end-to-end design that will be constructed for CTA. A number of prototype structures and cameras are being developed in order to investigate various alternative designs. In this contribution, these designs are presented, along with the technological solutions under study.}

\keywords{Cherenkov Telescope Array, SST, dual-mirror telescopes, SiPM, MAPM, structures and mirrors.}

\begin{document}
\maketitle

\section{Introduction}

The forthcoming Cherenkov Telescope Array (CTA) \cite{bib:Acharya, bib:Actis}, with its innovative approach based on the use of three different sizes of telescopes, will obtain a one-order-of-magnitude improvement with respect to the current Cherenkov telescope performance (achieved within HESS, MAGIC and VERITAS). CTA aims to provide global coverage of the sky from two observatory sites: a Southern array, implemented in particular for the exploration of both the Galactic plane and the extragalactic sky, and a Northern array, mainly devoted to the study of extragalactic sources. 
In the atmospheric showers originated by a gamma--ray primary, the Cherenkov light intensity is almost proportional to the gamma--ray energy (see e.g. \cite{bib:Mishev}). In general, large--diameter mirrors are needed to trigger low energy gamma-rays, while small mirrors are sufficient enough to trigger high energy events. Moreover, due to the very low gamma-ray fluxes at high energy, future Cherenkov telescopes like CTA must be able to catch events reaching the ground very far ($\sim 300-500$ m) from the telescope position, thus achieving effective areas of the order of  $10^6$ m$^2$; to trigger far showers, imaged at large off-axis angles, Cherenkov telescopes must be provided with sufficiently large fields of view. Optical dish diameters and field of view are the first parameters to be considered before analyzing other specific aspects such as optical design, mirror structure, focal plane sensors and electronics.
In this respect, due to the forward direction of Cherenkov light emission in air and the resulting Cherenkov light pool of fairly uniform illumination of about 200--250m diameter, for CTA an inter-telescope spacing of about 100m is needed at threshold energies to provide images in multiple telescopes. Well above this threshold, showers can be detected from outside the light pool, if the field of view is large enough. In CTA a small number (4 units for both northern and southern sites) of Large Size Telescopes (LST) of 23 m diameter will be deployed close to the centre of the array with $\simeq$100 m spacing. A larger number (up to 61 in the southern site, 15 in the northern site) of Medium Size Telescopes (MST) will cover a larger area, with an inter-telescope spacing of $\simeq$ 150 m. ÊThe southern site will include at least 25 single-mirror telescopes of 12 m and up to 36 Schwarzschild-Couder Telescopes of 9.5 m diameter. Above a few TeV the Cherenkov light intensity is such that showers can be detected even well outside the light pool by telescopes significantly smaller than the MST. To achieve the required sensitivity at high energies, a very large area on the ground needs to be covered by the Small Size Telescopes with a resolution of $\simeq$ 0.1$^\circ$ and a field of view of $\simeq$ 10$^\circ$. The SST sub--array can therefore be accomplished by 70 telescopes with a mirror area of 5--10 m$^2$ and $\simeq$ 300 m spacing, distributed across an area of 10 km$^2$ and within a radius of about 3 km. \\
The SST array will be implemented just on the southern site for reason of costs and taking into account that the very high energy emission can be observed just for galactic sources, unless non standard processes are invoked. 
In order to allow the implementation of the large number ($\sim70$) of telescopes foreseen for the SST sub-array, containing the costs to be $\le$ 500 kEuro per unit is essential. It should be noted that classical parabolic or Davies-Cotton (DC) single--mirror configurations have been used so far for Cherenkov telescopes, and they are adopted also for the CTA LST and MST telescopes respectively. However they are dominated by the cost of the camera, which is based on classical large-size photo-multipliers and they do not seem ideal for making the wide-field SST units. A possible solution is to implement the classical DC solution with Winston cone light guides, to squeeze the light with an aggressive concentration ratio, but with the crucial drawbacks of a difficult implementation and limited number of pixels \cite{bib:Modersky}.
But a particularly attractive solution to realize the SST telescopes is the use of a dual-mirror (2M) solution, adopting the so called Schwarzschild--Couder (SC) configuration. This enables good angular resolution across the entire field of view, almost 10$^\circ$ in diameter, and also reduces the effectiveÊ focal length and camera size \cite{bib:Vassiliev}. As has previously been demonstrated, 2M SC telescopes allow better correction of aberrations at large field angles and hence the construction of telescopes with a smaller focal ratio. This implies that, for a given primary mirror and angular pixel size, the physical pixels are smaller. This approach allows the use of low-cost, compact, and low power consumption cameras, based on silicon photomultipliers (SiPM) or multi-anode photomultipliers (MAPM), commercially available sensors with typical pixel size of $\sim 6$ $\times$ 6 mm$^2$. The trigger threshold of 1 TeV implies primary mirror diameter D $\simeq$ 4 m. The SST group within CTA has designed 2M telescopes which have the potential to provide the required optical performance and allow exploitation of these technologies.\\ 
Two SC structure designs for SST are currently being pursued, the ASTRI (Italy) and SST-GATE (France/UK). Designs of telescopes and cameras for both options are progressing well. The GATE and ASTRI groups will produce a joint 2M design for CTA based on results of prototype tests and recommendations of internal and external advisory committees. A common 2M camera will be produced by ASTRI/CHEC groups following prototype tests. Array operation with this telescope and camera combination will be tested at the CTA site.  
In the following, the optics and mechanical structure of the telescopes are described. A further section focuses on the two cameras. 

\section{The opto-mechanical designs}

\begin{figure}[!ht]
  \centering
  \includegraphics[width=0.4\textwidth]{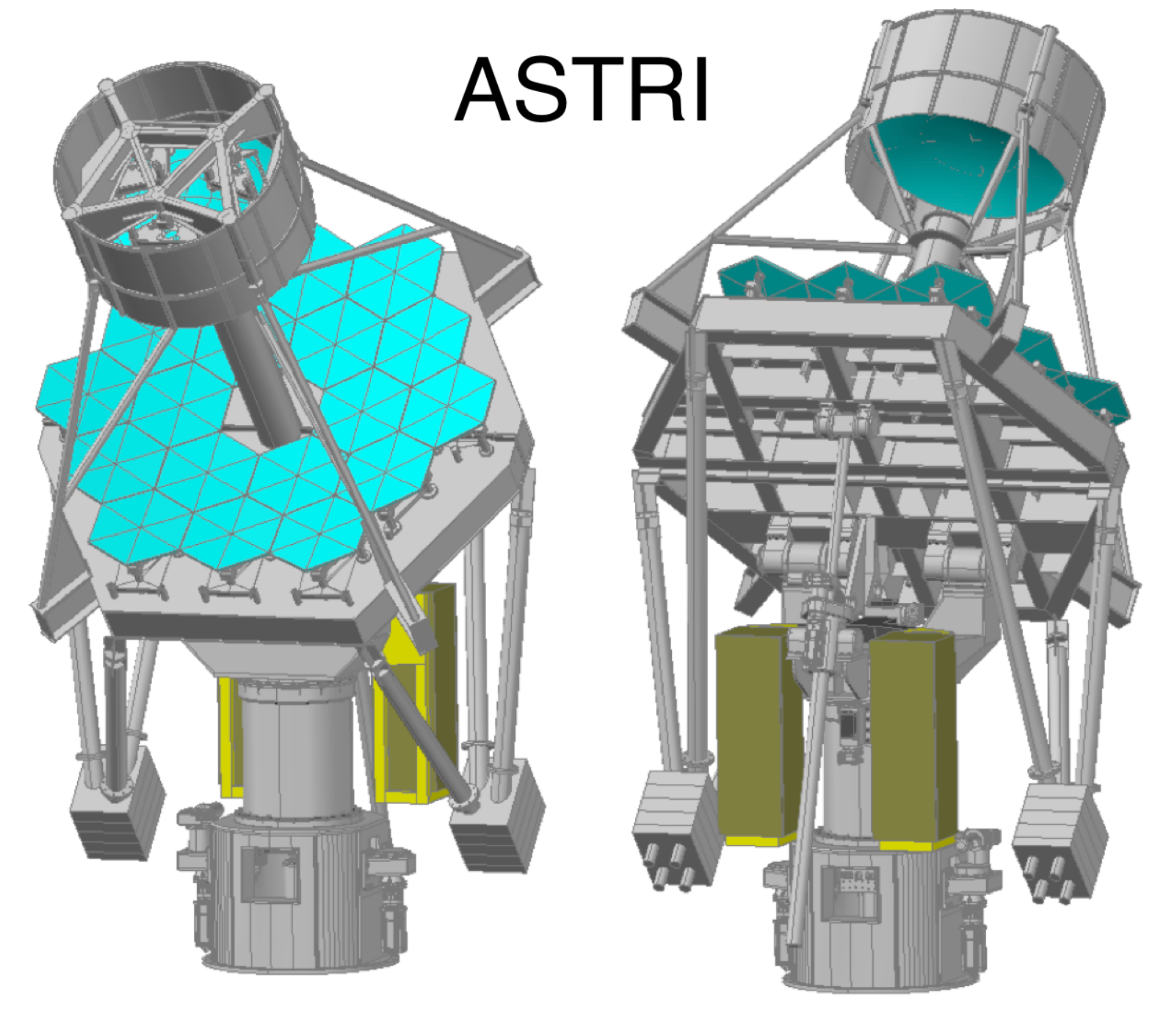}
   \includegraphics[width=0.4\textwidth]{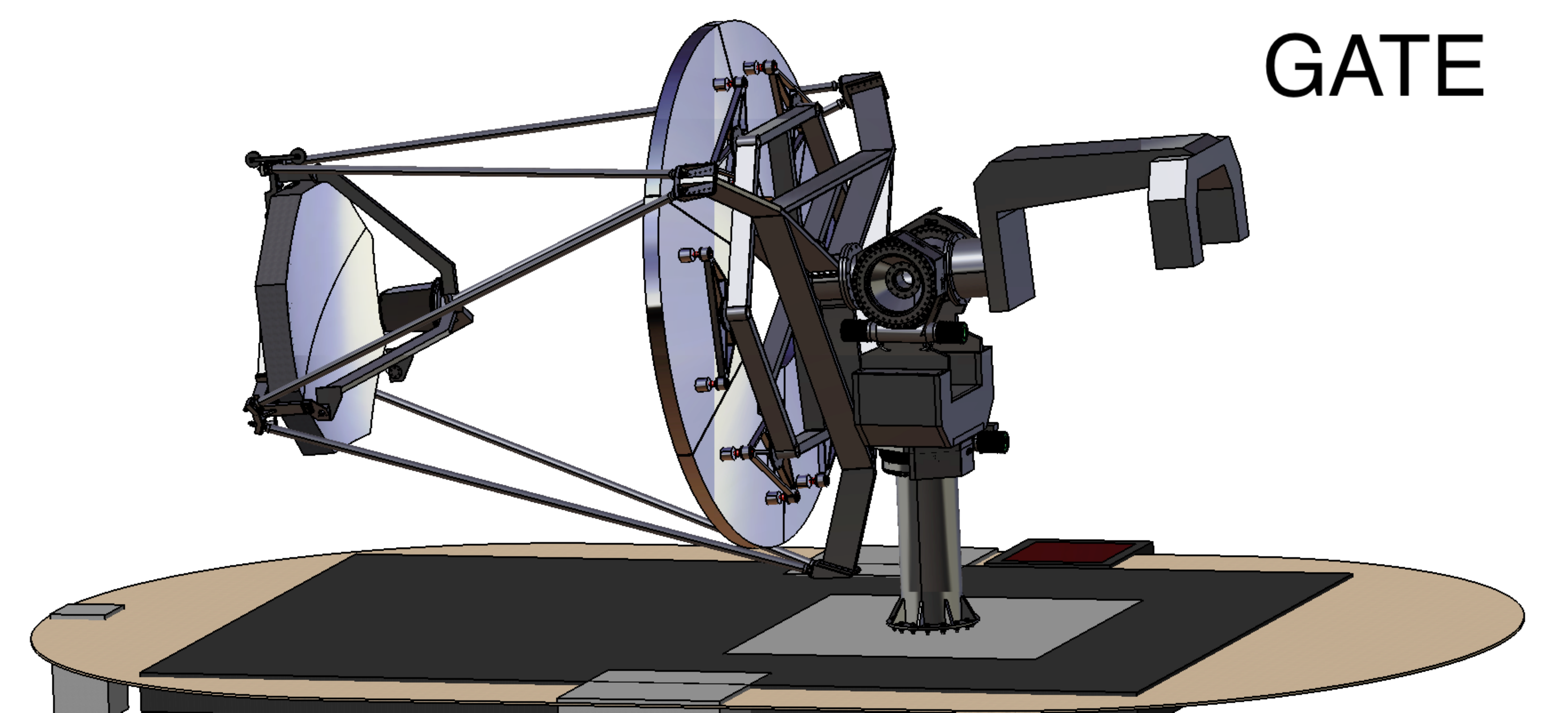}
  \caption{General view of the ASTRI and GATE telescope structures and electro-mechanical subsystems.}\label{fig:tel-str}
 \end{figure}

The Figure \ref{fig:tel-str} shows the 3-dimensional drawings of the telescope structures under development by the ASTRI and GATE SST collaborations. Matching the physical size of the pixels offered by MAPM or SiPM sensors (a few millimetres) to the required angular pixel size of the SST implies that the focal length of the telescope F is $\simeq$2 m. Ensuring sufficient collection area to obtain efficient triggering in the SST energy range, that is, a primary mirror of diameter about 4 m, then requires that the telescope's focal ratio be about 0.5. The proposed designs adopt the 
SC configuration ensuring a light collection efficiency higher than 80\% within the dimension of the pixels over a wide field. The mirror profiles are aspheric with substantial deviations from the main spherical component. The ASTRI and GATE designs differ somewhat a bit one to each other; hereafter we describe the ASTRI design as example.  The optical system has a plate scale of 37.5 mm/$^\circ$, an angular pixel size of approximately 0.17$^\circ$ and an equivalent focal length of 2150 mm. This setup delivers a corrected field of view up to 9.6$^\circ$ in diameter. Concerning the throughput, a mean value of the effective area of about 6 m$^2$ is achieved, taking into account: the segmentation of the primary mirror, the obscuration of the secondary mirror, the obscuration of the camera, the reflectivity of the optical surfaces as a function of the wavelength and incident angle, the losses due to the camera's protection window and finally the efficiency of the detector as a function of the incident angles (ranging from 25$^\circ$ to 72$^\circ$). The resulting telescope is compact, having a primary mirror (M1) diameter of 4 m and a primary-to-secondary distance of 3 m.
\\
Both the ASTRI and GATE telescopes adopt an alt-azimuthal design in which the azimuth axis will permit a rotation range of $\pm$270$^\circ$. The mirror dish is mounted on the azimuth fork which allows rotation around the elevation axis from -5$^\circ$ to +95$^\circ$. 
The mast structure that supports the secondary mirror and the camera has different solutions between the two projects. In order to balance the torque due to the overhang of the optical tube assembly with respect to the horizontal rotation axis, counterweights are also envisaged. For the ASTRI telescope it is proposed to construct the primary mirror as a set of 18 hexagonal-shaped panels having 850 mm face-to-face dimension. Three different types of mirror profiles are necessary to reproduce the M1 profile. GATE will exploit petal-shaped segments made in polished Aluminum. 
For both telescopes, the aim is to build a monolithic secondary mirror.
\\
\subsection{Structure designs}
\textbf{The ASTRI design.}\\
The top of the column is interfaced to the azimuth fork, this interface being the azimuth bearing. The fork supports the elevation assembly (dish, counterweight and the quadrupod that is the structure supporting the secondary mirror and connected to the dish) through a linear actuator and bearings at its upper ends. The fork is composed of welded steel box sections. The M1 dish structure is attached to the azimuth fork using two preloaded tapered roller bearings, one on each of the fork's arms. The dish is a ribbed steel plate of about 40 cm thickness. To this are attached the supports for the mirror segments, each of which includes a single and a double axis actuator and a bearing, which allows steering of the segments for alignment purposes. The eccentric quadrupod legs counteract the lateral deformations of the mast structure due to gravitational and wind loads, while the central tube increases the torsional stiffness of the structure. At the upper end of the mast is located the structure for supporting the monolithic secondary mirror. This consists of three actuators attached to the mirror via whiffletrees to ensure the load is spread over a sufficiently large area. In addition, three lateral arms support the transverse components of the secondary mirror's weight as they vary with the orientation of the telescope. All structures are made of steel and protected against corrosion by paint. The azimuth drive is located at the base of the column and is composed of two pinions, driven by electric motors, that couple with a rim gear. Axially pre-loaded ball bearings complete the azimuth assembly. The linear actuator, which drives the elevation, consists of a preloaded ball screw driven through a gearbox by an electric motor. The orientation of the telescope is determined using absolute encoders located on each of the azimuth and altitude axes.  Finite Element Analysis (FEA) has been used to evaluate the performance of the system. The lowest frequency eigenmode of the oscillations of the structure is 4.5 Hz. FEA has also been carried out to determine the effects on the telescope of temperature gradients. More details on the ASTRI design are presented by Canestrari et al. in \cite{bib:canestrari}.\\ 
\noindent
\textbf{The GATE design}\\
The design philosophy was to split the telescope into functions so that each one becomes as independent as possible to alleviate the constraints on the design. The GATE structure is based on a shallow fork structure for which one counterweight is needed, mounted internally to the fork. The azimuth is mounted on the tower and the fork holds the elevation. The boss-head connects the elevation subsystems to both the counterweight and the optical structure. The mast and truss structure is composed of the support of the M1 dish, tubes in a Serrurier-like hexapod configuration, the support of the M2 dish and arms holding the camera. Aluminum and carbon fiber have also been considered, but finally steel has been selected to optimize costs and easiness of manufacturing and mounting. The alt-azimuth system is the same for the azimuth and elevation and is composed of slewing bearing with worm gear (1 for azimuth, 2 for 
elevation), motors and absolute encoders. FEA has determined that the lowest frequency eigenmodes of the oscillations of the telescope involve transverse motion of the secondary with respect to the primary and are about 5 Hz. Torsional eigenmodes of oscillation have lowest frequencies of about 12 Hz. These values refer to a preliminary design not yet optimized. More details are presented by Zech et al. in \cite{bib:zech}.\\ 
%
\section{Cameras}
\begin{figure}[!ht]
  \centering
  \includegraphics[width=0.32\textwidth]{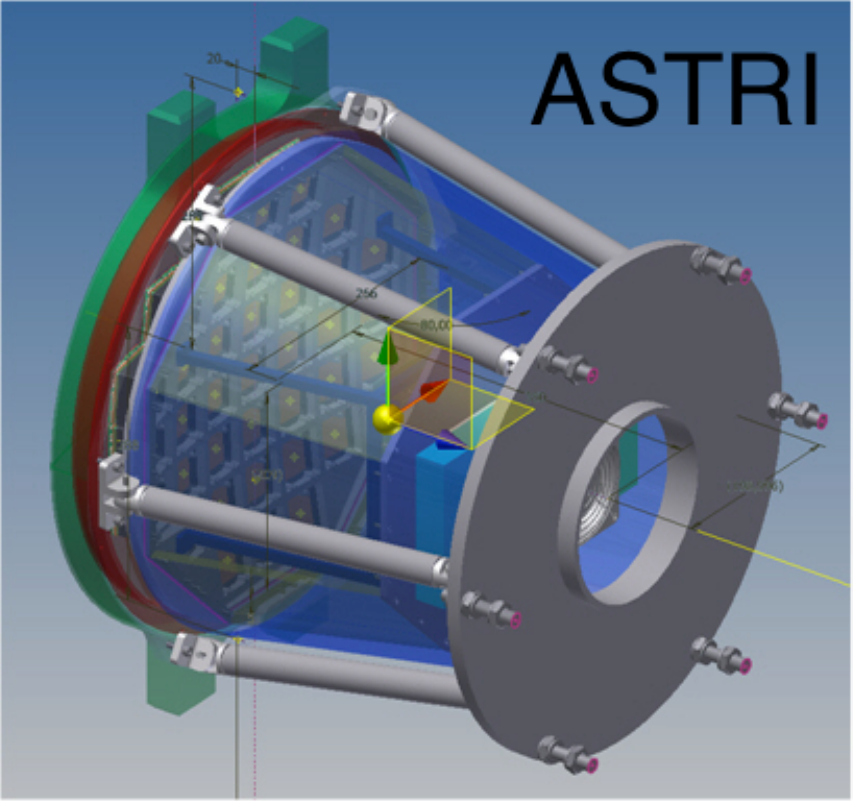}
   \includegraphics[width=0.32\textwidth]{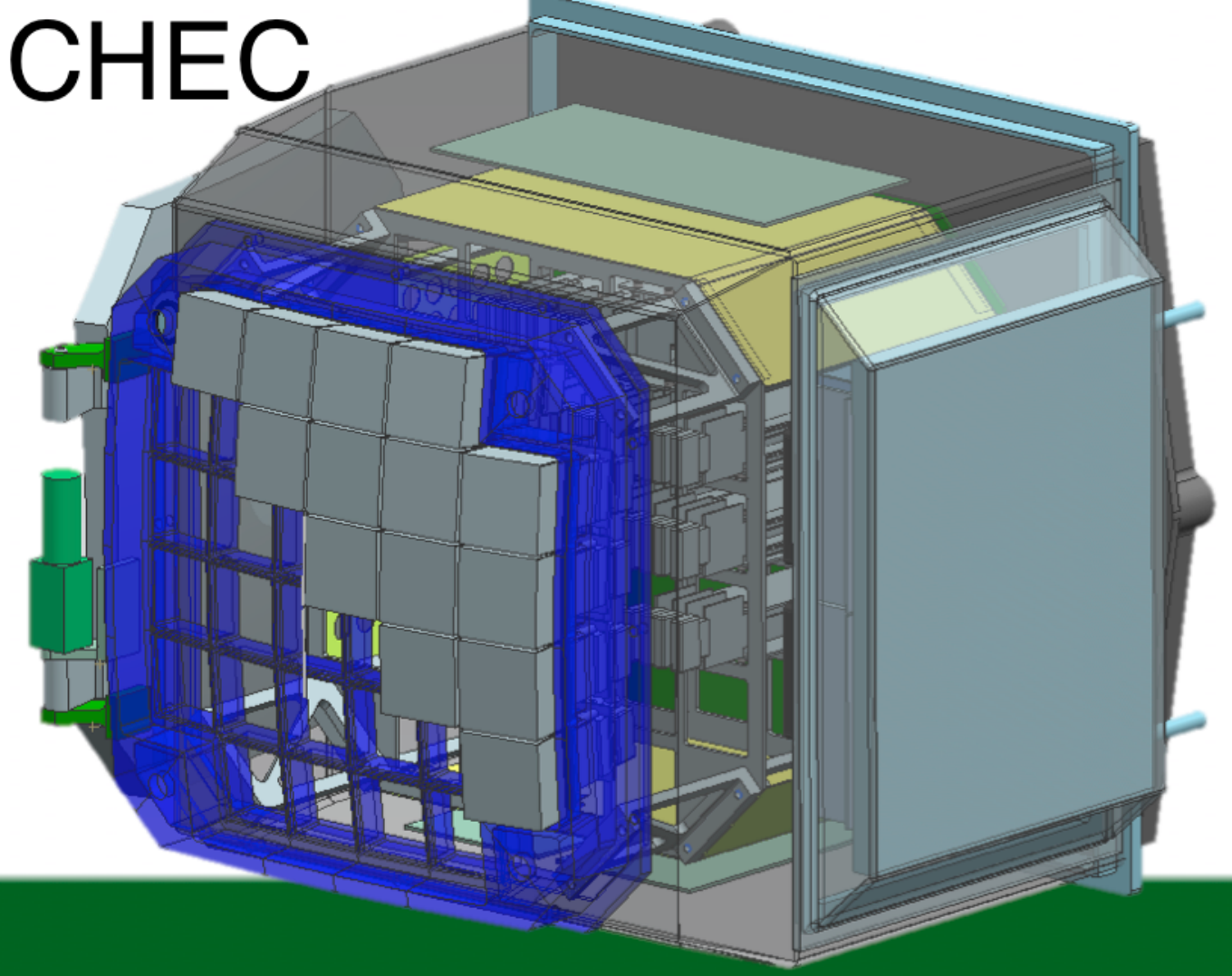}
  \caption{General view of the ASTRI and CHEC cameras.}
  \label{fig:camstr}
 \end{figure}

Figure \ref{fig:camstr} shows the 3-dimensional drawings of the camera structures under development by the ASTRI and CHEC collaborations. Details are hereafter reported.\\
\noindent
\textbf{The ASTRI design.}\\
The ASTRI camera \cite{bib:Catalano} adopts SiPM as photosensors. Each pixel contains 3600 microcells, each being an avalanche photodiode operated in quenched Geiger mode. These cells are 50$\times$50 $\mu$m$^2$ wide, giving a fill factor of 70\%. The particular device chosen is the Hamamatsu S11828-334 monolithic multi-pixel SiPM, consisting of 4x4 pixels of roughly 3x3 mm$^2$ each. Four Hamamatsu pixels are grouped together to form one camera pixel with a physical size 6.2$\times$6.2 mm$^2$, matching the required angular size. Four of the Hamamatsu devices are put together to form a unit. Four such units then form a module called a Photon Detection Module (PDM). This module is composed of 16 Hamamatsu devices and has dimensions of $56\times 56$ mm$^2$. The PDM are constructed by plugging the Hamamatsu devices into connectors attached to a printed circuit board (PCB). Under each unit on the PCB there is a small temperature sensor, allowing the temperature of the SiPM to be monitored, providing a route through which the temperature dependent SiPM gain can be stabilized. 
The  Front End Electronics (FEE) boards supply the power for the SiPM, perform the readout and form the first trigger signals. The EASIROC (Extended Analogue SiPM Readout Chip) ASIC will be adopted. A range of simulation tools has been produced to aid the design of the readout and trigger. With the presently available Hamamatsu SiPM devices there are small gaps between the sensors when they are mounted to form units and PDMs. Truncated pyramidal light guides, made of high refractive index glass (2.5 mm thick) to guarantee a large acceptance angle, are finally implemented in order to reduce the light losses due to dead areas. 
In addition to the FEE mentioned above, design of the Back-End Electronics (BEE) is underway. This will use a Field Programmable Gate Array (FPGA) and local memory to provide interfaces to the CTA data acquisition, to the camera controller and to the CTA clock. There will also be circuitry to provide the various DC voltages needed to power the elements of the camera. 
A preliminary design of the chassis of the complete camera is also shown in figure \ref{fig:camstr} (upper panel). The total height of the camera is about 30 cm. The camera lid can be seen in this picture.  This subsystem can be closed to protect the sensors from the elements. As the focal plane of the 2M telescopes is convex, with a radius of curvature of 1 m, the PDM must be attached to a precisely machined curved plate. Below the sensor plane is the support structure to which further electronics boards and the cooling system can be attached.
\noindent
\textbf{The CHEC camera.}\\
The CHEC collaboration will develop two cameras, one based on MAPM and the other on SiPM sensors. The first prototype will be the one based on MAPM sensors \cite{bib:White} and is hereafter described. The default photosensor for the Compact High-Energy Camera is the Hamamatsu H10966 MAPM. This consists of 64 pixels each of size 6$\times6$ mm$^2$, in a unit of dimensions $52\times52$ mm$^2$. Thirty-two MAPM can be used to cover the focal plane, providing a field of view of about 9$^\circ$.
Each FEE module provides the high voltage supply needed by a MAPM, samples the signals produced by its 64 channels at a frequency of about 500 MHz, forms a first level trigger by applying thresholds to sums of four pixels and outputs a digitized waveform for each of the MAPM channels. CHEC will make use ofÊ modules designed at SLAC both for the SC MSTs and for CHEC based on the TARGET ASIC. Minor modifications to this are needed and these will be carried out in collaboration with SLAC. The TARGET modules \cite{bib:Becthol}, as they are, cannot be connected directly to the MAPM in the CHEC because the curvature of the focal plane would then require large gaps between the MAPM. The MAPM is instead attached to a preamplifier board and then a structure containing a twisted length of ribbon cable, allowing bending in two planes, which carries the signals from the MAPM to the electronics. The preamplifier board behind the MAPM allows the PMT to be operated at low gain (typically 10$^5$), important given the high counting rate they will experience due to background photons. The preamplifier also allows shaping of the MAPM signal and hence optimization of the performance of the FEE. Further, if it becomes clear that SiPM will offer better performance per unit cost than MAPM, the MAPM photosensor plane with its preamplifiers can be replaced with a SiPM-based system with new preamplifiers ensuring the correct signal shape enters the TARGET module. A mechanical frame that provides the required rigid support has been designed. 
Prototype mechanical structures have shown that this system functions as hoped, allowing the MAPM to be placed on a curved surface.
The trigger signals provided by the FEE modules must be combined and examined to select candidate Cherenkov events against the night-sky background. This camera trigger system forms part of the BEE, which is also responsible for processing data from the FEE modules and distributing clock signals with the required level of precision to the FEE modules. Recent developments mean that it is feasible to process trigger signals with nanosecond accuracy and sub-nanosecond delay correction from all FEE modules in one or two FPGA (Field Programmable Gate Array) devices. The mechanical aspects of the camera include a support matrix for the MAPM, internal elements for support of the electronics, a cooling system and an external structure, which includes a lid and the interface to the telescope. The MAPM support matrix must ensure precise focal-plane positioning. The internal structure must allow adequate cooling of the electronics and remain stable on decade timescales under repeated camera movements. Thermal modeling of the camera will be done during the camera design to assess the cooling and control requirements, which will be implemented during the mechanical prototyping. The external structure must be weather-proof and minimize dust ingress, and provide minimal additional shadowing of the primary mirror. 

\section{Conclusions}

The dual mirror telescopes and their cameras are being developed by three consortia, ASTRI (Italy/INAF), GATE (France/Paris Observ.) and CHEC (universities in UK, US and Japan) which are merging their efforts in order to finalize an end-to-end design that will be constructed for CTA. A number of prototype structures and cameras are being developed in order to investigate various alternative designs and a tradeoff study will be performed to determine the final design.


\vspace*{0.5cm}

\footnotesize{{\bf Acknowledgment: }{We gratefully acknowledge financial support from the following agencies and organizations: Ministerio de Ciencia, Tecnolog\'{i}a e Innovaci\'{o}n Productiva (MinCyT), Comisi\'{o}n Nacional de Energ\'{i}a At\'{o}mica (CNEA), Consejo Nacional de Investigaciones Cient\'{i}ficas y T\'{e}cnicas (CONICET), Argentina; State Committee of Science of Armenia, Armenia; FAPESP (Funda\c{c}\~{a}o de Amparo \`{a} Pesquisa do Estado de S\~{a}o Paulo, Brasil; Ministry of Education, Youth and Sports, MEYS LE13012, 7AMB12AR013, Czech Republic; Ministry of Higher Education and Research, CNRS-INSU and CNRS-IN2P3, CEA-Irfu, ANR, Regional Council Ile de France, Labex ENIGMASS, OSUG2020 and OCEVU, France; Max Planck Society, BMBF, DESY, Helmholtz Association, Germany; Istituto Nazionale di Astrofisica (INAF), MIUR, Italy; ICRR, The University of Tokyo, JSPS, Japan; Netherlands Research School for Astronomy (NOVA), Netherlands Organization for Scientific Research (NWO), Netherlands; The Bergen Research Foundation, Norway; Ministry of Science and Higher Education, the National Centre for Research and Development and the National Science Centre, Poland; MINECO support through the National R+D+I, CDTI funding plans and the CPAN and MultiDark Consolider-Ingenio 2010 programme, Spain; Swedish Research Council, Royal Swedish Academy of Sciences, Sweden; Swiss National Science Foundation (SNSF), Switzerland; Durham University, Leverhulme Trust, Liverpool University, University of Leicester, Royal Society, Science and Technologies Facilities Council, UK; U.S. National Science Foundation, U.S. Department of Energy, Argonne National Laboratory, Barnard College, University of California, University of Chicago, Columbia University, Georgia Institute of Technology, Institute for Nuclear and Particle Astrophysics (INPAC-MRPI program), Iowa State University, Washington University McDonnell Center for the Space Sciences, USA.}}

\end{document}